\documentstyle[prb,aps,preprint,tighten]{revtex}

\begin{document}

\title{Transport properties of a quantum wire
in the presence of impurities and long-range Coulomb forces}
\author{H. Maurey and T. Giamarchi}
\address{Laboratoire de Physique des Solides, Universit{\'e} Paris--Sud,
                   B{\^a}t. 510, 91405 Orsay, France\cite{junk}}
\maketitle

\begin{abstract}
One-dimensional electron systems
interacting with long-range Coulomb forces (quantum wires)
show a Wigner crystal structure.
We investigate in this paper the transport properties of such a Wigner
crystal
in the presence of impurities. Contrary to what happens when only
short-range interactions are included, the system is dominated by $4
k_F$ scattering on the impurities.
There are two important length scales in such
a problem: one is the pinning length above which the (quasi-)long-range
order of the Wigner crystal is destroyed by disorder. The other length
$\xi_{cr}$ is the length below which Coulomb interactions are not
important and the system is behaving as a standard Luttinger liquid with
short-range interactions. We obtain the frequency and
temperature dependence of the conductivity. We show that such a system
is very similar to a classical charge density wave pinned by impurities,
but with important differences due to quantum fluctuations and long-range
Coulomb interactions. Finally we discuss our results in comparison
with experimental systems.
\end{abstract}

\section{Introduction}

Recently nanostructure technology has made it possible to create
quasi one-dimensional electronic structures, the so-called quantum
wires \cite{field_1d,scott-thomas_1d,kastner_coulomb,goni_gas1d,%
calleja_gas1d,tarucha_wire_1d}.
Experimentally situations have been reached where the width of such
a wire is of the order of the Fermi wavelength of the conduction electrons,
which makes it a good realization of a one-dimensional electron gas
\cite{calleja_gas1d}.

In such a system one expects the electron-electron interactions to play an
important role. In particular, at variance with what happens in other
one-dimensional conductors, the long-range Coulomb interaction in a quantum
wire is not screened since the wire contains only one channel of
electrons. One can therefore expect very different physical properties
than those of a Luttinger liquid with short-range interactions.
It has been proposed that due to these long-range interactions, the electrons
in a quantum wire will form a Wigner crystal \cite{schulz_wigner_1d}.
The formation of a Wigner crystal can be described as a modulation of the
charge density $\rho(x) \sim \rho_0 \cos(Qx+2\sqrt{2}\Phi)$ where
$\rho_0$
is the uniform amplitude of the charge density, $Q= 4 k_F$ its wave
vector
and $\Phi$ describes the location and the motion of the Wigner crystal.
The existence of such a Wigner crystal should have observable consequences on
the transport properties of the system. Indeed, in the presence of impurities
the Wigner crystal will be pinned: the phase $\Phi(x)$ adjusts to the impurity
potential on a scale given by $L_0$ called the pinning length.
This process of pinning is analogous to what happens in charge density
waves in the presence of impurities \cite{lee_rice_cdw,fukuyama_pinning}.
Since, in the presence of long-range Coulomb interactions,
the most divergent fluctuation is now a $4 k_F$ density
fluctuation \cite{schulz_wigner_1d}, the transport properties are
dominated by $4 k_F$ scattering on impurities, and not the
usual $2 k_F$ scattering, as was assumed previously
\cite{ogata_wires_2kf,fukuyama_wires_2kf}. Due to the $4 k_F$
scattering, one can also expect different
transport properties than those of a Luttinger liquid with short-range
interactions
\cite{apel_impurity_1d,suzumura_scha,giamarchi_loc_lettre,giamarchi_loc}
where $2 k_F$ fluctuations are the dominant one.

Non linear $I-V$ curves have been observed experimentally
which could be interpreted as the result of pinning
\cite{kastner_coulomb}. Up to now only short wires have been made,
for which only few impurities are in the wire and dominate the transport.
Even in that case what has mainly been focussed on theoretically
is a system with short-range interactions
\cite{kane_qwires_tunnel_lettre,kane_qwires_tunnel,glazman_single_impurity}.
For long wires, it is important to consider the case of a
uniform disorder, e.g. a thermodynamic number of impurities, as well as
the long-range Coulomb forces.

In this
paper we study the effects of disorder on the transport
properties of a quantum wire.
Although the problem is very close to that
of a charge density wave pinned by impurities, there are important
differences that are worth investigating. Due to the long-range
 nature of the forces one can expect modifications of the pinning
length and frequency dependence of the conductivity. In addition quantum
fluctuations have to be taken into account, and for the case of
short-range forces are known to drastically affect the transport properties
compared to a classical situation \cite{suzumura_scha,giamarchi_loc}.

The plan of the paper is as follows. The model is derived in
section~\ref{model}. Effects of the pinning and the pinning length are
studied in section~\ref{static}. The frequency dependence of the
conductivity is computed in section~\ref{conductivity} and the
temperature dependence of the conductivity (and conductance)
is discussed in section~\ref{temperature}.
Discussion of
the comparison with experiments and conclusions can be found in
section~\ref{conclusion}. Some technical details about the treatment
of quantum fluctuations can be found in the appendices.

\section{Model} \label{model}

We consider a gas of electrons confined in a channel of
length $L$, with a width
$d\ll L$ and a thickness $e\ll d\ll L$.
We will assume that both $d$ and $e$ are small enough for
the system to be regarded as one-dimensional, meaning that
only one band is filled in the energy-spectrum of the electrons.
Such a situation will be realized when $d$ and $e$ become comparable to
the Fermi wavelength.
In the following we will therefore keep only the degrees of freedom
along the wire.
Since we are interested only in low energy excitations
we can linearize the spectrum around the Fermi points
and take for the free part of the Hamiltonian:
\begin{equation} \label{free}
H_0 = v_F \sum _k (k-k_F) a^{\dag }_{+,k}a_{+,k} +  (-k-k_F)
a^{\dag }_{-,k}a_{-,k}
\end{equation}
where $v_F$ is the Fermi velocity and
$a^{\dag }_{+,k}$ ($a^{\dag }_{-,k}$) is the creation
operator of an electron on the right(left)-going branch
with wave-vector $k$. In addition we assume that the electrons
interact through the Coulomb interaction
\begin{equation}\label{interact}
H_c = \frac1{2}\int _0 ^L \int _0 ^L dx dx' V(x-x') \rho (x) \rho (x')
    =  \frac1{2L} \sum _k V_k \rho _k \rho _{-k}
\end{equation}
In a strictly one-dimensional theory, a $\frac1{r}$ Coulomb potential
has no Fourier transform because of the divergence for
$r\to 0$. In the real system such a divergence does not exist
owing to the finite width d of the wire.
We will use for $V(r)$ the
following approximate form \cite{gold_1dplasmon} which cuts the
singularity at $r \approx d$, and gives the correct asymptotic behavior
at large $r$
\begin{equation}
V(r) = \frac{e^2}{\sqrt{r^2+d^2}}
\end{equation}
the Fourier transform of which is
\begin{equation}
V(q)=\int_{-L/2}^{L/2} dr V(r)e^{iqr} \approx 2e^2 K_0(qd)
\end{equation}
where $K_0$ is a Bessel function, and one has assumed the wire to be
long enough $L\to \infty$.
In the following we shall frequently use the asymptotic expression
\begin{equation}
K_0(qd) \approx -\ln (qd) \qquad \text{when} \qquad qd \ll 1
\end{equation}

The model (\ref{free}) plus (\ref{interact}) has been studied by
Schulz \cite{schulz_wigner_1d} who showed that
the system is dominated by $4k_F$ charge
density wave fluctuations, which decay as
\begin{equation}
\langle \rho_{4k_f}(x)\rho_{4k_f}(0)\rangle \sim e^{-\ln^{1/2}(x)}
\end{equation}
The presence of such a $4k_F$ charge fluctuation can be
viewed as the formation of a Wigner crystal.
In order to describe the pinning of such a Wigner crystal
we add to the hamiltonian (\ref{free}) and (\ref{interact})
the contribution due to impurities.
We assume that impurities are located in the wire at random sites
$X_j$, and that each impurity acts on the electrons with a
potential $V_{imp}$.
We will assume in the following that the potential
due to the impurities is short-ranged, and will replace it by a delta
function.
\begin{equation}
V_{imp}(x-X_j) = V_0\delta (x-X_j)
\end{equation}
The part of the Hamiltonian stemming from a particular configuration of
the impurities is then
\begin{equation} \label{imp}
H_{imp} = \sum_j \int_0^L V_0\delta (x-X_j) \rho (x) = \sum_j V_0\rho (X_j)
\end{equation}

In order to treat the problem we use the representation of fermion
operators in term of boson operators
\cite{solyom_revue_1d,emery_revue_1d}.
One introduces the phase field
\begin{equation}
\Phi (x) = - {i\pi \over L} \sum_{k\ne 0} {1 \over k}e^{-ikx}(\rho _{+,k}+
\rho _{-,k})
\end{equation}
where $\rho _{+,k}$($\rho _{-,k}$) are the charge density operators for
right(left)-moving electrons,
and $\Pi$, the  momentum density conjugate to $\Phi$.
The boson form for (\ref{free}) plus (\ref{interact})
is \cite{schulz_wigner_1d}
\begin{equation} \label{starting}
H_0+H_c = {u \over 2\pi }\int_0^L dx \lbrack K(\pi \Pi)^2
        + {1 \over  K} (\partial _x \Phi)^2 \rbrack
        + {1 \over \pi^2} \int _0^L \int _0^L dxdx' V(x-x')
           (\partial _x \Phi(x)) (\partial _{x}\Phi(x'))
\end{equation}
$K$ is a number containing the backscattering effects due to the
Fourier components of the interaction close to $2 k_F$ and $u$ is the
renormalized Fermi velocity due to the same interactions
\cite{solyom_revue_1d,emery_revue_1d,schulz_wigner_1d}.
We have taken $\hbar =1$ in (\ref{starting}).
The long-range nature of the Coulomb interaction manifests itself in
the last term of (\ref{starting}). As we shall precise in the following,
both $K$ and the Coulomb potential $V$ control the strength of quantum
effects.

Since for (\ref{starting}) the most divergent fluctuation
corresponds to a $4 k_F$ charge modulation \cite{schulz_wigner_1d},
we will consider only
the coupling of the impurities with this mode and ignore the $2 k_F$
part of the charge density. The range of validity of such an
approximation will be discussed in the following.
Using the boson representation of the density
\cite{solyom_revue_1d,emery_revue_1d} and impurity Hamiltonian
(\ref{imp}), the total Hamiltonian becomes
\begin{eqnarray}
H & = & {u \over 2\pi }\int _0^L dx \lbrack K(\pi \Pi)^2
  + {1 \over  K} (\partial _x \Phi )^2 \rbrack
  + \sum _j V_0\rho _0 \cos(4 k_F X_j + 2\sqrt{2}\Phi (X_j))
  \nonumber \\
  &   &  + \frac1{\pi ^2} \int _0^L \int _0^L dxdx'
V(x-x') (\partial _x\Phi(x)) (\partial _{x}\Phi(x')) \label{total}
\end{eqnarray}
where $\rho_0$ is the average density of electrons.
The hamiltonian (\ref{total}) has similarities with the phase
Hamiltonian of a pinned charge density wave
\cite{fukuyama_pinning}.
Similarly to the CDW case one can expect the
phase to distort to take advantage of the impurity potential, leading to
the pinning of the Wigner crystal.
As for standard CDW, one has to distinguish
between strong and weak pinning on the impurities
\cite{fukuyama_pinning}. In the first case
the phase adjusts itself on each impurity site. This corresponds to a
strong impurity potential or dilute impurities. In the weak pinning
case, the impurity potential is too weak or the impurities too close
for the phase to be able to adjust on each impurity site, due to the
cost in elastic energy. Although the problem has similarities with the
CDW problem, there are two important a priori physical differences that
have to be taken into account:
compared to the CDW case, one has
to take into account the long-range Coulomb interaction. One can
expect such an interaction to make the Wigner crystal more rigid than a CDW
and therefore more difficult to pin. In addition, for the Wigner
crystal, one cannot neglect the quantum term
($\Pi^2$) as is usually done for the CDW problem owing to the large
effective mass of the CDW. In the absence of long-range interactions such
a term is known to give important quantum corrections
\cite{suzumura_scha,giamarchi_loc} on both the pinning length and the
conductivity.

In the following sections we will examine both cases of strong and weak
pinning.

\section{Calculation of the pinning length}
\label{static}
Let us first compute the pinning
length $L_0$ over which the phase $\Phi (x)$ in the ground state varies
in order to take advantage of the impurity potential.
If the impurities are dilute enough, or the impurity potential strong
enough,
the phase $\Phi (x)$ adjusts on each impurity site such that
$\cos(4 k_F X_j + 2\sqrt{2}\Phi (X_j))=-1$. This is the so-called strong
pinning regime \cite{fukuyama_pinning} where
the pinning length is the distance between impurities
$L_0 = n_i^{-1}$. If the impurities are dense enough, or their potential
weak enough then the cost of elastic and Coulomb energy in distorting the
phase has to be balanced with the gain in potential energy. One is in
the weak pinning regime where the pinning length can be much larger than
the distance between impurities.
In this regime, we calculate $L_0$ using
Fukuyama and Lee's method developed for the CDW
\cite{fukuyama_pinning,lee_coulomb_cdw}.
This method neglects the quantum fluctuations of the phase, and the
effect of such fluctuations will be discussed at the end of this
section.
One assumes that the phase $\Phi$ varies on a scale $L_0$. One can
therefore divide the wire in segments of size $L_0$ where the phase is
roughly constant and takes the optimal value to gain the maximum pinning
energy. $L_0$ is determined by
optimizing the total gain in energy, equal to the gain in potential
energy minus the cost in elastic and Coulomb energy. If one assumes
that the phase varies of a quantity of order $2\pi$ over a length
$L_0$, the cost of elastic energy per unit length is
\begin{equation}\label{eps-el}
{\cal E}_{el} = {u \over 2\pi K} {1 \over \alpha L_0^2}
\end{equation}
where $\alpha$ is a number of order unity depending on the precise
variation of the phase. Since the impurity potential varies randomly
in segments of length $L_0$, the gain per unit length due to pinning is
\cite{fukuyama_pinning}
\begin{equation} \label{eps-imp}
{\cal E}_{\rm imp}(L_0) = - V_0\rho _0 ({n_i \over L_0})^{1 \over 2}
\end{equation}
In our case we also have to consider the cost in Coulomb energy.
\begin{equation}
{\cal E}_{coul} = {1 \over L} {1 \over \pi ^2} \int _0^L dx \int _0^L dx'
V(x-x') \langle \partial _x \Phi \rangle_{av} \langle \partial _{x'} \Phi
\rangle_{av}
\end{equation}
where the subscript {\it av} indicates that the quantity is averaged over all
impurity configurations. Since one assumes
that the phase varies of a quantity of order $2\pi$ over a length
$L_0$, the phases for  electrons distant of more than
$L_0$ are uncorrelated, so that the interactions between such pairs of
electrons do not contribute to the energy.
The calculation can thus be reduced to the evaluation of the energy for
a segment of length $L_0$
\begin{equation} \label{eps-coul}
{\cal E}_{coul} \approx {1 \over \pi ^2}{1 \over L_0}
L_0 \int _{-L_0}^{L_0} du V(u) {\langle \Phi ^2(x) \rangle _{av} \over L_0^2}
=  {2e^2 \over \pi ^2\alpha L_0^2} \ln {L_0 \over d}
\end{equation}
where $\alpha$ is the constant introduced in (\ref{eps-el}).

The minimization of the total energy provides a self-consistent expression
for $L_0$:
\begin{equation} \label{implicit}
L_0 = ({8e^2 \over \alpha \pi ^2 V_0\rho _0n_i^{1 \over 2}})^{2 \over 3}
\ln ^{2 \over 3} ({CL_0 \over d})
\end{equation}
where $C$ is a constant of order one
\begin{equation} \label{constant}
C = e^{({\pi u \over 4Ke^2} - {1 \over 2})}
\end{equation}
Taking typical values $u=3\times 10^7cm.s^{-1}$ so that
$\hbar u=3.15\times 10^{-20}$  e.s.u. and $K=0.5$,
one gets
$C \approx 0.75$. For these typical values of the parameters
the contribution of the elastic (short-range) part of the hamiltonian to
that result is negligible compared to that of the Coulomb term.
In the following, since we expect $\frac{L_0}{d} \gg 1$ we approximate
$\ln \frac{CL_0}{d} \sim \ln \frac{L_0}{d}$.

Neglecting $\log(\log)$ corrections, one can solve (\ref{implicit})
to get
\begin{equation} \label{length}
L_0 = \Biggl(\frac{8e^2}{\alpha \pi ^2 V_0\rho_0
       n_i^{\frac1{2}}}\Biggr)^{\frac2{3}}
\ln ^{\frac2{3}} \Biggl( \frac 1{d} \Bigl({8e^2 \over \alpha
\pi ^2V_0\rho _0n_i^{1/2}}\Bigr)^{\frac2{3}}\Biggr)
\end{equation}
Compared to the pinning length of a CDW
\cite{fukuyama_pinning},
$L_0 \approx (\frac {v_F}{\alpha \pi V_0\rho_0n_i^{1/2}})^{2/3}$, the
pinning length (\ref{length}), is enhanced by a logarithmic factor.
This is due to the Coulomb interaction which enhances the rigidity of
the system and makes it more difficult to pin than a classical
CDW.

The expression (\ref{length}) has been derived for the weak pinning case
where $L_0 \gg n_i^{-1}$. The crossover to the strong pinning regime
occurs when the
phase can adjust itself on each impurity
site and $L_0 = n_i^{-1}$.
One can introduce a dimensionless quantity $\epsilon_0$
characterizing the two regimes
\begin{equation} \label{epsilono}
\epsilon_0 = \frac{\alpha \pi^2 V_0\rho _0}{8 n_i e^2}
\ln ^{-1} \Biggl( \frac 1{d} \Bigl({8e^2 \over \alpha
\pi ^2V_0\rho _0n_i^{1/2}}\Bigr)^{\frac2{3}}\Biggr)
\end{equation}
The weak pinning corresponds to $\epsilon_0 \ll 1$, and the strong
starts at $\epsilon_0 \simeq 1$. Compared to a CDW where
$\epsilon_0 = \frac{V_0\rho_0}{n_iv_f}$, the domain of weak pinning is
larger due to the Coulomb interaction. This is again a consequence of
the enhanced rigidity of the system that makes it more difficult to pin.
To study the conductivity it is also convenient to introduce
\begin{equation} \label{epsilon}
\epsilon = \frac{V_0 \rho _0}{n_ie^2}
\end{equation}
Indeed we have evaluated, using typical values $d=10^{-8}$m and
$L_0 \simeq 10^{-6}$m, (estimated for typical wires in
section~\ref{conclusion}),
that $\epsilon_0 \simeq 2 \epsilon$ so that $\epsilon$ can also be used
as criterion to distinguish the two regimes of pinning.

Expressions (\ref{length}) and (\ref{epsilono}) do not take into account
the effects of quantum fluctuations. In the absence of Coulomb
interactions, the quantum fluctuations drastically increase the
pinning length compared to the classical case
\cite{suzumura_scha,giamarchi_loc} giving a pinning length (for a $4
k_F$ dominant scattering)
\begin{equation}
L_0 \sim (1/V_0)^{2/(3-4K)}
\end{equation}
To compute the effect of the quantum fluctuations in the presence of the
Coulomb interaction we use the self-consistent harmonic approximation
\cite{suzumura_scha} for the cosine term in (\ref{starting})
\begin{equation} \label{scha}
\cos (Q x+2\sqrt{2}(\Phi_{cl}+\hat{\Phi}))=e^{-4\langle \hat{\Phi}^2(x)
\rangle}
\cos(Qx+2\sqrt{2}\Phi_{cl}) (1-4(\hat{\Phi}^2(x)-\langle \hat{\Phi}^2(x)
\rangle))
\end{equation}
where $\Phi =\Phi_{cl}+\hat{\Phi}$ and $\hat{\Phi}$ represents
the quantum fluctuations around the classical solution $\Phi_{cl}$.
The average $\langle\hat{\Phi} \rangle$ has to be done self
consistently. Such a calculation is performed in appendix~\ref{quant}
and one obtains for the pinning length
\begin{equation}
L_0 =({8e^2 \over \alpha \pi ^2 V_0\rho _0\gamma n_i^{1 \over 2}})^{2 \over 3}
\ln ^{2 \over 3} ({CL_0 \over d})
\end{equation}
where $\gamma = e^{-4\langle\hat{\Phi}^2\rangle} \approx e^{-\frac{8\tilde{K}}
{\sqrt{3}}\ln^{1/2}V_0}$ and
$\tilde{K} = \frac{\sqrt{\pi uK}}{2\sqrt{2}e}$ instead of
(\ref{length}). The quantum fluctuations can thus
be taken into account by replacing $V_0$ by the effective impurity
potential $V_0 \gamma$. There is an increase of the pinning length
due to the quantum fluctuations which can be considerable since
$\gamma \ll 1$. Opposite to what happens for the case of short
range interactions, there is no correction in the exponent for the
pinning length. This can be traced back to the fact that the correlation
functions decay much more slowly ($e^{-\ln^{1/2}(r)}$ instead of a
power law), therefore the system is much more ordered and
the fluctuations around the ground state are much less important. As a
consequence even if one is dealing with a system of electrons, and not a
classical CDW, the
Coulomb interactions push the system to the classical limit
where quantum fluctuations can be neglected except for the redefinition
of the impurity potential $V_0 \to V_0 \gamma$. Note that this effect
can be very important quantitatively, since $L_0$ is very large for
dilute impurities. Such a fluctuation effect also contributes to make
the system more likely to be in the weak pinning regime.

We will in the following make the assumption that all quantum
fluctuation effects have been absorbed in the proper redefinition of the
pinning length. Such an approximation will be valid as long as one is
dealing with properties at low enough frequencies. At high frequencies
the effect of quantum fluctuations will again be important and will be
examined in section~\ref{LargeFreq}.

\section{Calculation of the conductivity}
\label{conductivity}

In order to study the transport properties, one makes an expansion
around the static solution $\Phi_0(x)$ studied in section (\ref{static})
that minimizes the total energy \cite{fukuyama_pinning}, assuming
that the deviations $\Psi(x,t)$ are small
\begin{equation}
\Psi (x,t) = \Phi (x,t) -\Phi _0(x)
\end{equation}
One can expand the Hamiltonian  in $\Psi(x,t)$ to quadratic order
\begin{eqnarray} \label{expansion}
{\cal H} _{\Psi } & = & {u \over 2\pi } \int _0^L dx K(\pi \Pi )^2 + {1
\over K}(\partial _x\Psi )^2
+  {1 \over \pi ^2} \int _0^L \int _0^L dx dx' V(x-x') \partial _x\Psi
\partial _{x'}\Psi \nonumber \\
 & & - 4\sum _j V_0\rho _0 \cos(4k_F X_j +2\sqrt{2}\Phi _0(X_j))
(\Psi (X_j))^2
\end{eqnarray}
This expansion is valid in the classical case. We assume that for the
quantum problem all quantum corrections are absorbed in the proper
redefinition of the pinning length $L_0$, as explained in
section~\ref{static} and appendix~\ref{quant}.
Such corrections do not affect
the frequency dependence of the conductivity. From
Kubo formula and the representation of the current in terms
of the field $\Psi$, the conductivity takes the form
\begin{equation}
\sigma (\omega ) = 2i\omega ({e \over \pi} )^2 {\cal D}
(0,0;i\omega _n) \rfloor _{i\omega _n \rightarrow \omega + i0^+}
\end{equation}
where ${\cal D}(q,q';i\omega _n)$ is the Green's function of the field
$\Phi$
\begin{equation}
{\cal D}(q,q';i\omega _n) = \int _0^{\beta} d\tau e^{i\omega
_n\tau} \langle T_{\tau }\Psi _q(\tau )\Psi _{-q'}(0)\rangle
\end{equation}
with $\Psi _q(\tau )=e^{H\tau }\Psi _qe^{-H\tau }$, $\beta = T^{-1} (k_B=1)$
and $\omega _n=2\pi nT$, where
$T$ is the temperature, and $T_ {\tau }$ is the time-ordering operator.
Our problem is then reduced to the evaluation of this Green function. From
(\ref{expansion}) one gets the Dyson equation
\begin{equation} \label{dysondep}
{\cal D}(q,q';i\omega _n) = {\cal D}_0(q,i\omega _n) \lbrack \delta _{q,q'}
+ 8V_0\rho _0 \sum _{q''} S(q''-q){\cal D}(q'',q';i\omega _n) \rbrack
\end{equation}
where
\begin{equation}
S(q)=\frac1{L} \sum _j e^{iqX_j} \cos(QX_j+2\sqrt{2}\Phi _0(X_j))
\end{equation}
After averaging over all impurity configurations (\ref{dysondep})
becomes
\begin{equation}
\langle {\cal D}(q,q';i\omega _n)\rangle_{av} =
\delta _{q,q'}{\cal D}(q,i\omega _n)
= {1 \over {\cal D}_0(q,i\omega _n)^{-1}-\Sigma (q,i\omega _n)}
\end{equation}
where the self-energy term $\Sigma $ contains all connected contributions to
${\cal D}$, and ${\cal D}_0$ is the free Green Function
\begin{equation}
{\cal D}_0(q,i\omega _n)=
{\pi uK \over \omega _n^2 + q^2u^2(1+{2KV(q) \over \pi u})}
\end{equation}
In a similar fashion than for CDW we will compute the self-energy, using
a self-consistent Born approximation \cite{fukuyama_pinning}, for the
two limiting cases of strong and weak pinning.

\subsection{Weak pinning case $(\epsilon \ll 1)$}
\label{weak}

In that case, as for standard CDW \cite{fukuyama_pinning},
the self-energy can be expanded to second-order in perturbation,
$\Sigma  \approx \Sigma _1 + \Sigma _2$.
Indeed we easily verify that in the weak pinning case
$\Sigma _1 \sim\Sigma _2 \sim n_i^2(V_0\rho_0/n_i)^{4/3}$,
whereas for $n \ge 1$, $\Sigma _{2n+1}=0$ and
$\Sigma _{2n} \sim n_i^2(\frac{V_0\rho_0}{n_i})^{\frac{2+2n}{3}}$. Since
$\frac{V_0\rho_0}{n_i} \sim \epsilon e^2 \ll 1$, self-energy terms of higher
order than $\Sigma_2$ are negligible. $\Sigma_1$ is
easily computed as
\begin{equation}  \label{sigma1}
\Sigma _1 = 8V_0\rho_0\langle S(0)\rangle_{av} =
-8V_0\rho _0 ({n_i \over L_0})^{1 \over 2}
\end{equation}
since again one can divide the wire into $L/L_0$ segments of length $L_0$,
and use, as for equation (\ref{eps-imp}), the random-walk argument of
reference \onlinecite{fukuyama_pinning} which gives
\begin{equation}  \label{cos}
\frac1{L}\langle\sum_j \cos(QX_j + 2\sqrt{2}\Phi _0(X_j))\rangle_{av}
\approx \sqrt{\frac{n_i}{L_0}}
\end{equation}

$\Sigma _2$ is given by
\begin{equation}
\Sigma _2 = (8V_0\rho _0)^2 \sum _{q''} {\cal D} _0(q'',i\omega
_n)\langle S(q''-q)S(q-q'') \rangle_{av}
\end{equation}
If one assumes that there is
no interference between scattering on different impurities (single site
approximation), then
the exponentials in $\langle S(q''-q)S(q-q'')\rangle_{av}$
cancel and we find
\begin{eqnarray} \label{sigma2}
\Sigma_2 & = & ({8V_0\rho _0 \over L})^2 \sum _{q''} {\cal D}
_0(q'',i\omega _n)
\langle \sum _j \cos^2(QX_j + 2\sqrt{2}\Phi _0(X_j))\rangle_{av}  \\
& =&  64{n_i \over 2}(V_0\rho _0)^2 {1 \over L} \sum _{q''} {\cal D}
_0(q'',i\omega _n) \nonumber
\end{eqnarray}
The approximation
$\frac1{L}\langle \sum _j \cos^2(QX_j + 2\sqrt{2}\Phi _0(X_j))\rangle_{av}
\approx \frac1{2}n_i$ is valid in the weak pinning case only. A more general
result is
\begin{eqnarray}
\frac1{L}\langle \sum _j \cos^2(QX_j + 2\sqrt{2}\Phi _0(X_j))\rangle_{av}
& = & \frac1{L}\langle \sum_j \frac1{2}(1+\cos 2(QX_j + 2\sqrt{2}\Phi _0(X_j))
\rangle_{av} \nonumber \\
& \approx & \frac1{2}(n_i + \sqrt{\frac{n_i}{L_0}}) \label{SquCos}
\end{eqnarray}
but in the weak pinning
case it can be simplified using $n_iL_0 \gg 1$.

$\Sigma _2$ given by (\ref{sigma2}) diverges as
$\frac1{|\omega_n|}\ln \frac1{|\omega_n|}$ when $|\omega_n| \to 0$, so
one has to compute $\Sigma$ self-consistently, and
replace ${\cal D}_0$ by ${\cal D}$ in the calculation of $\Sigma_2$.
(\ref{sigma2}) is replaced by
\begin{equation} \label{prime}
\Sigma _2' = 32n_i(V_0\rho _0)^2  {1 \over L} \sum _{q''}{\cal
D}(q'',i\omega _n)
= 32n_i(V_0\rho _0)^2  {1 \over L} \sum _{q''}\lbrack {\cal
D}_0^{-1}(q'',i\omega _n) - \Sigma (q'',i\omega _n) \rbrack ^{-1}
\end{equation}
giving the self-consistent equation for $\Sigma$
\begin{equation} \label{exacte}
\Sigma = \Sigma_1 + \Sigma _2'
       = -8V_0\rho_0 \sqrt{\frac{n_i}{L_0}}
         + 16 n_i(V_0\rho_0)^2(\pi uK)\frac 2{\pi}
 \int _0^{\infty} \frac{dq}{-\omega^2-\pi uK\Sigma +
                       q^2u^2(1+\alpha_c K_0(qd))}
\end{equation}
where we have done the analytic continuation $i\omega_n \to \omega
+i0^+$ and we have noted $\alpha_c = \frac{4Ke^2}{\pi u}$.
It is convenient to rescale (\ref{exacte}) by the pinning frequency
$\omega^*$ defined by
\begin{equation} \label{PinFreq}
\omega^{*3}\ln^{1/2}\frac{\tilde{u}}{d\omega^*}
 =16n_iu(V_0\rho_0\pi K)^2 \frac 1{\sqrt{\alpha_c}}
={8\pi ^{5/2}(uK)^{3/2} \over e} n_i(V_0\rho_0)^2
\end{equation}
where $\tilde{u}=u\sqrt{\alpha_c}$.
In terms of $L_0$, (\ref{PinFreq}) can be rewritten as
\begin{equation}
\omega^* \ln^{1/6}\frac{\tilde{u}}{\omega^*d}=
 4\alpha^{-2/3}\tilde{u}L_0^{-1}\ln^{2/3}\Bigl({L_0 \over d}\Bigr)
\end{equation}
Neglecting $\log(\log)$ factors, and in the limit $L_O \gg d$ allowing
to discard the constants in the logarithm
($\ln\frac{\tilde{u}}{\omega^*d} \approx \ln \frac {L_0}{d}$)
we obtain
\begin{equation} \label{pinningob}
\omega^*\approx 4\alpha^{-2/3}\tilde{u}L_0^{-1}\ln^{1/2}\Bigl({L_0 \over d}
\Bigr)
\end{equation}
Leaving aside a factor $4\alpha^{-2/3}$, $\omega^*$ given by
(\ref{pinningob}) is
the characteristic frequency of a segment of the wire of length $L_0$.
Indeed if we modelize the wire as a collection of independent oscillators of
typical length $L_0$ and use the dispersion law $\omega \sim
q\ln^{1/2}q$ of
the Wigner Crystal \cite{schulz_wigner_1d}, those oscillators have
the frequency $\omega_0=\tilde{u}L_0^{-1}\ln^{1/2}\Bigl({L_0 \over d}\Bigr)$.
Numerically we find $4\alpha^{-2/3} \approx 1$ so that actually
$\omega^* \approx \omega_0$.
Introducing the rescaled quantities
$y=\frac{\omega}{\omega^* }$ and $G={\pi uK\Sigma \over \omega^{*2}}$,
we rewrite (\ref{exacte}) as
\begin{equation} \label{rescaled}
G = G_1 + G'_2
\end{equation}
with $G_1 = -\alpha^{\frac 1{3}}$
and
\begin{equation} \label{integrale}
G'_2=\ln^{1/2}\frac{\tilde{u}}{d\omega^*}\sqrt{\alpha_c}\frac 2{\pi}
\int_0^{\infty} \frac{dt}{-y^2-G+t^2(1+\alpha_cK_0(\frac{\omega^*d}{u}t))}
\end{equation}
The rescaled conductivity is:
\begin{equation} \label{conduct}
\omega^* \Re e \sigma (y) = - \frac{2uKe^2}{\pi} y\Im m ({1 \over -y^2-G})
\end{equation}

The full solution of (\ref{rescaled}) has to be obtained numerically,
but it is possible to obtain analytically the asymptotic expressions at
small and large frequencies. To evaluate the integral $G'_2$, one notices
that there is a frequency $\omega_{cr}$ above which the Coulomb term
will be negligible compared to the kinetic (short-range) term.
$\omega_{cr}$ defines a crossover length $\xi_{cr} \sim u/\omega_{cr}$
which is roughly given by
\begin{equation}
\xi_{cr} \sim d e^{1/\alpha_c} = d e^{\frac{\pi u}{4 K e^2}}
\end{equation}
Using a numerical estimate for $\alpha_c$ and the values of the Bessel
function one gets $\alpha_c K_0(x) \sim 1$ for $x\sim 1.5$, giving a
crossover frequency  $\omega_{cr}
\sim 1.5\frac {\tilde{u}}{d} \sim 10^{14} Hz$. Such a frequency
is two order of magnitude larger than the pinning frequency
$\omega^*$.
For frequencies above
$\omega_{cr}$ the system is dominated by short-range interactions:
in that case the dominant fluctuations are always the $2k_F$ charge
fluctuations and not the $4k_F$ ones, and therefore  the model
(\ref{total}) is not applicable. One has to take into account
the pinning on a $2k_F$ fluctuation as done in reference
\onlinecite{suzumura_scha,giamarchi_loc}.
Note that it
makes sense to use a one-dimensional model to describe the behavior
above $\omega_{cr}$ only if $\xi_{cr} \gg d$. This can occur for example
if the short-range interactions are strong enough so that $K$ is very
small. With the numerical values of $u$ that seem relevant for
experimental quantum wires and assuming that $K$ is not too small
$K\sim 0.5$, one gets $\xi_{cr} \sim d$. Therefore one can assert that in
all the range of frequencies for which the problem can be considered
as one-dimensional, Coulomb interactions will dominate.
Consequently the result of the integration (\ref{integrale}) is,
when $\omega^*\sqrt{-y^2-G} \ll \omega_{cr}$
\begin{equation}  \label{small}
G'_2 =\frac 1{\sqrt{-y^2-G}}\ln^{-1/2}\frac{\tilde{u}}{d\omega^*\sqrt{-y^2-G}}
\ln^{1/2}\frac{\tilde{u}}{d\omega^*}
\end{equation}
Let us focus on small frequencies $\omega \ll \omega^*$.
We will show that in that limit
$\omega^*\sqrt{-y^2-G}\sim \omega^* \ll \omega_{cr}$, so that we use
(\ref{small}) and replace in (\ref{rescaled})
\begin{equation}
G = G_1 + {1 \over \sqrt{-y^2-G}} \ln ^{-1/2}
{\tilde{u} \over \omega^* d\sqrt{-y^2-G}}
\ln^{1/2}\frac{\tilde{u}}{d\omega^*}
\qquad \text{when} \qquad y \ll 1
\end{equation}
$G$ tends to a limit $G_0$ verifying
\begin{equation} \label{Gzero}
G_0 = G_1 + (-G_0)^{-1/2}\ln ^{-1/2}\Bigl(\frac{\tilde{u}}{d\omega^*}
(-G_0)^{-1/2}\Bigr)\ln^{1/2}\frac{\tilde{u}}{d\omega^*}
\end{equation}
Equation (\ref{Gzero}) has different classes of solutions depending on the
value of $\alpha$ (zero, one or two roots),
but the only physically relevant situation
is the case of a single solution
(for a discussion, see reference \onlinecite{fukuyama_pinning}). The
corresponding value of
$G_0$ is
\begin{equation} \label{value}
G_0 \approx -2^{-2/3}
\end{equation}
Expanding (\ref{Gzero}) in terms of $y$ and of $G-G_0$ around
that solution, we find
\begin{equation} \label{Gexpansion}
G-G_0 = \pm i\frac 2{\sqrt{3}}(-G_0)^{1/2}y
\qquad \text{for} \qquad y\ll 1
\end{equation}
We assumed in deriving (\ref{value}) and (\ref{Gexpansion}) that
$\ln^{-1}(\tilde{u}/d\omega^* )$ is small compared to $1$. This can be
verified numerically for the parameters we are taking. Using
$\omega^* \sim 1\times 10^{12}Hz$ (as estimated in section
\ref{conclusion})  and taking $d \sim 10^{-8}m$ one obtains
$\ln^{-1}(\frac {\tilde{u}}{d\omega^*}) \sim 0.2$.
Replacing in(\ref{conduct}), we find for the conductivity
\begin{equation} \label{result}
\omega^* \Re e \sigma (y) = {uKe^2 \over \pi } \frac 8{\sqrt{3}}y^2
\qquad (y \to 0)
\end{equation}

One can now check that the hypothesis
$\omega^*\sqrt{-y^2-G}\sim \omega^* \ll \omega_{cr}$
is indeed verified.
$\sqrt{-y^2-G}$ is well defined for $y \ll 1$
since $-G_0$ is positive, and $\sqrt{-y^2-G} \sim 1$ since
$\sqrt{-G_0}$ is of the order of $1$.

We have plotted in figure~\ref{conducti}, the full frequency behavior
of the conductivity, together with the analytic estimate at small
frequencies.
The small $\omega$ behavior as well as the general shape of the
conductivity is very similar to the one of a classical charge density
wave: the small $\omega$ conductivity is behaving as $\omega^2$, there
is a maximum at the pinning frequency $\omega^*$ followed by a decrease
in $1/\omega^4$. As shown in appendix~\ref{quant} the quantum
fluctuations do not change the frequency dependence for frequencies
lower than
$\omega^*$. The large frequency behavior will be analyzed in details in
section~\ref{LargeFreq}.

The low frequency conductivity obtained in our approximation is to be
contrasted with the previous result of Efros and
Shklovskii \cite{efros_coulomb_gap} who find
that the low frequency conductivity of a one-dimensional electron gas in
the presence of Coulomb interactions should behave as $\omega$.
This result is derived in a very different physical limit where the
localization length is much smaller than the interparticle distance,
whereas the implicit assumption to derive the model (\ref{total}) is that the
localization length is much larger than the interparticle distance
$k_F^{-1}$. In the limit that was considered in
\onlinecite{efros_coulomb_gap} the phase
$\phi$ would consist of a series of kinks of width $l$ the localization
length and located at random positions (with an average spacing
$k_F^{-1} \gg l$).
The low-energy excitations that are taken into account in
\onlinecite{shklovskii_conductivity_coulomb}, would
correspond to soliton-like
excitations for the phase
$\phi$, where the phase jumps by $2\pi$ between two distant kinks.
In the physical limit we are considering $k_F^{-1} \ll L_0$, the phase
$\phi$ has no kink-like structure but rather smooth distortions between
random values at a scale of order $L_0$. To get the dynamics, the
approximation we are using only retains the small ``phonon'' like
displacements of the phase $\phi$ relative to the equilibrium position
and no ``soliton'' like excitations are taken into account.
In the absence of Coulomb interactions the phonon-like excitations
alone, when treated exactly in the classical limit $K\to 0$
are known \cite{vinokur_cdw_exact} to give the
correct frequency dependence of the conductivity
$\omega^2\ln^2(1/\omega)$ (the self-consistent Born approximation only
gets the $\omega^2$ and misses the log correction).
When Coulomb
interactions are included and one is in the limit where the localization
length is much larger than the interparticle distance, it is not clear
whether soliton-like interactions similar to those considered by Efros
and Shklovskii have to be taken into
account. From the solution of a uniform sine-Gordon equation, one
could naively say that solitons are only important when the quantum
effects are large $K \sim 1$. In the classical limit $K \to 0$, the
phonon modes have a much lower energy than the soliton excitations, and
the physical behavior of the system should be dominated by such modes.
We would therefore argue that the conductivity is given correctly by
our result (up to possible log corrections) and to behave
in $\omega^2$, and not
$\omega$, at least if the system is classical enough ($K$ small) thanks
to the {\bf short-range}  part of the interaction. If our assumption is
correct the crossover towards the Efros and Shklovskii result when the
disorder becomes stronger would be very interesting to study.

\subsection{Strong pinning case $\epsilon > 1$}

Let us now look at the other limit case of strong pinning.
In that case one cannot expand the self-energy $\Sigma$,
all the
single-site contributions \cite{fukuyama_pinning} have to be summed.
The result of that summation is
\begin{equation}
\Sigma = (-8V_0\rho_0n_i) {1 \over 1+8V_0\rho_0A}
\end{equation}
where $A$ is defined by
\begin{equation}\label{self-consist}
A=\frac1{L} \sum_{q} {\cal D}(q,i\omega_n)
\end{equation}
Here we rescale the conductivity by the characteristic frequency
\begin{equation}
\omega_0= n_i \tilde{u}\ln ^{1/2}({1 \over d n_i})
\end{equation}
corresponding to a pinning length $L_0 \sim n_i^{-1}$. It is thus the analog
of $\omega^*$, to a factor $4\alpha^{-2/3} \approx 1$.
We use as rescaled parameters
$\overline{y}={\omega \over \omega_0}$
and $\overline{G}={\pi uK\Sigma \over \omega_0^2}$, in which terms
the expression of the conductivity is similar to
(\ref{conduct}), where we replace $y$,$G$ and $\omega^*$ respectively by
$\overline{y}$, $\overline{G}$, and $\omega_0$.
The resolution is quite similar to what was done for the CDW
\cite{fukuyama_pinning}, so that we give only the main results.

The exact equation  on the rescaled self-energy $\overline{G}$ is
\begin{equation} \label{self}
\overline{G}=-\lbrack \frac1{2\pi^2}\ln\frac{\tilde{u}}{\omega_0d}
\frac1{\epsilon}+ \sqrt{\alpha_c}\ln^{1/2}\frac{\tilde{u}}{\omega_0d}
\frac1{\pi} \int_0^{\infty}
\frac{dt}{-(-y^2-\overline{G})+t^2(1+\alpha_cK_0(\frac{\omega_0d}{u}t))}
\rbrack ^{-1}
\end{equation}
where $\epsilon$, strength of the pinning, was defined in (\ref{epsilon}).
The numerical resolution of this equation gives the conductivity plotted
on Fig.~\ref{conductiStrong}, for different values of $\epsilon$.
There is a gap below a frequency $\omega_{\text{lim}} < \omega_0$
close to the pinning frequency and tending to it as $\epsilon$ gets
bigger. In the extremely strong pinning limit $\epsilon \gg 1$ one
can obtain analytically the conductivity. The equation for the self
energy (\ref{self}), after replacement of the integral by its analytical
approximate which can be taken from (\ref{small}) since
$\omega_0 \ll \omega_{cr}$, is:
\begin{equation}
 \overline{G}= -2\Bigl(\ln^{-1/2}\frac{\tilde{u}}{\omega_0d}\Bigr)
 \sqrt{-y^2-\overline{G}}
 \ln^{1/2}\frac{\tilde{u}}{\omega_0d\sqrt{-y^2-\overline{G}}}
\end{equation}
where the integral is given by (\ref{weak})
since $\omega_0 \ll \omega_{cr}$.
In this limit $\omega_{lim}=\omega_0$
and the conductivity is given near the threshold by
\begin{equation}
\omega_0 {\cal R}e\sigma(\overline{y}) \approx \frac{4\sqrt{2}uKe^2}{\pi}
\sqrt{\overline{y}-1}
\end{equation}
This gap below $\omega_{\text{lim}}$ is not physical and is an
artifact of considering only the mean distance $n_i^{-1}$
between impurities.
In the real system there is a finite probability of finding neighboring
impurities farther apart than $n_i^{-1}$. Such configurations
will give contributions at frequencies smaller than
$\omega_{\text{lim}}$.
An estimation of those contributions can be done in a similar way than
for a CDW \cite{gorkov_cdw_strong,fukuyama_pinning}.
The probability of finding two neighboring impurities at a distance $l$
is $n_ie^{-n_i l}$. In the strong pinning case where we model our pinned
CDW by a collection of independent oscillators with frequencies
${u\pi \over l}\ln^{1/2}(l/d)$,
the conductivity for $\omega < \omega_{\text{lim}}$ will then be
proportional to the sum of the contributions over all possible $l$
\begin{eqnarray}
{\cal R}e\sigma(\omega) & \sim & \int_0^{\infty} dl \; n_i e^{-n_i l}
\delta(\omega -{u\pi \over l}ln^{1/2}{l \over \pi d})\nonumber \\
& \sim & \omega ^{-2}\ln^{1/2}\frac1{\omega d}
e^{-\pi n_i \frac u{\omega}\ln^{1/2}\frac u{\omega d}}
\end{eqnarray}
Compared to a CDW, the conductivity in the pseudo gap is
lowered in the presence of Coulomb interactions. This can again be
related to the fact that the long-range forces make the Wigner crystal
more rigid.

\subsection{Large frequency conductivity}
\label{LargeFreq}

We focus now on large frequencies $\omega \gg \omega^*$,($\omega_0$),
where we expect the physics to be determined over segments of typical size
$l_{\omega} \sim \frac{\tilde{u}}{\omega} \ll L_0$,($n_i^{-1}$),
so that intuitively the behavior of the conductivity should be
independent of whether we are in the strong or weak pinning regime.
And indeed at high $\omega$ the conductivity can always be computed
using the approximation $\Sigma  \approx \Sigma _1 + \Sigma _2$,
whatever the pinning is, since the self-energy terms $\Sigma_n$'s are
of order $(\frac1{\omega})^{n-1}$.
But we recall that we made drastic assumptions on the phase $\Phi$,
depending on the pinning regime. To be consistent, they should
give similar results at high frequencies.
Let's start first from a weak pinning regime: at $\omega \ll \omega^*$ we
supposed the physics to be determined on domains of length $L_0$ on which
the phase $\Phi$ is roughly constant. If we now increase $\omega$ above
$\omega^*$ we simply replace $L_0$ by $l_{\omega}$ in the evaluation
of $\Sigma_1$ and $\Sigma_2$. More precisely (\ref{cos}) and (\ref{SquCos})
are replaced by:
\begin{eqnarray}
\frac1{L}\langle \sum_j \cos(QX_j +2\sqrt{2}\Phi_0(X_j)) \rangle_{av} &=&
\sqrt{\frac{n_i}{l_{\omega}}} \\
\frac1{L}\langle \sum_j \cos^2(QX_j+2\sqrt{2}\Phi_0(X_j))\rangle_{av}
&=&\frac1{2}n_i(1+\frac 1{\sqrt{n_il_{\omega}}})
\end{eqnarray}
This is valid of course as long as $l_{\omega} \gg n_i^{-1}$, above which
those averages saturate at values:
\begin{eqnarray} \label{satur}
\frac1{L}\langle \sum_j\cos(QX_j+2\sqrt{2}\Phi_0(X_j)) \rangle_{av} &=& n_i \\
\label{satur2}
\frac1{L}\langle \sum_j\cos^2(QX_j+2\sqrt{2}\Phi_0(X_j))\rangle_{av} &=& n_i
\end{eqnarray}
Starting from the strong pinning case and keeping the picture of the phase
being adjusted on each impurity site we find expressions identical to
(\ref{satur}) and(\ref{satur2}), regardless of the frequency.

In the end, using results of section \ref{weak} we compute the conductivity
 to be
\begin{equation} \label{highcon}
\omega^* \Re e \sigma (y) = \frac{c_{\Phi}4uKe^2}{\pi}
y^{-4}\ln^{-1/2}\frac{\tilde{u}}{d\omega^*y}\ln^{1/2}\frac{\tilde{u}}
{d\omega^*}
\end{equation}
when $\omega^*,\omega_0 \ll \omega \ll \omega_{cr}$, and where $c_{\Phi}$
is a numerical coefficient between $\frac1{2}$ and $1$. More
precisely
\begin{eqnarray} \label{cphi}
c_{\Phi} &=& \frac1{2}(1+\frac1{\sqrt{n_il_{\omega}}}) \qquad
\text{for} \qquad l_{\omega} \ge n_i^{-1} \\
     c_\Phi    &=& 1 \qquad \text{for} \qquad l_{\omega} \le n_i^{-1}
\nonumber
\end{eqnarray}
which sums up both weak and strong pinning results.

The result (\ref{highcon}) does not take into account the effect of
quantum fluctuations. Such effects are expected to become important for
frequencies larger than the pinning frequency.
For short-range interactions, using renormalization group techniques
\cite{giamarchi_loc,giamarchi_umklapp_1d}, one can show that if it is
possible to
neglect the renormalization of the interactions by disorder (for example
for very weak disorder) the conductivity becomes
(for a $4 k_F$ pinning)
$\sigma(\omega) \sim \omega^{4K-4}$ instead of $\omega^{-4}$ due to
quantum effects, and would be $\sigma(\omega) \sim \omega^{K-3}$
for $2 k_F$ scattering.
Although one can derive these results
and get the conductivity at high frequency for long-range interactions,
using the memory function formalism \cite{gotze_fonction_memoire}
in a way similar to
\onlinecite{giamarchi_umklapp_1d,giamarchi_attract_1d},
we will show here how to use the SCHA to get the high frequency
conductivity.
A naive way to take the frequency into account in the SCHA
is to divide the system into segments of length $\frac u{\omega}$, and
look at the system on scale of such a segment. Using this method it is
possible to rederive the results for the short-range interactions
and tackle the case of long-range interactions in which we
are interested. Such a calculation is performed
in appendix~\ref{HighFreq}. Instead of (\ref{highcon}), one gets
\begin{equation} \label{highconq}
\omega^* \Re e \sigma (y) = \frac{c_{\Phi} 4uKe^2}{\pi}
y^{-4}\ln^{-1/2}\frac{\tilde{u}}{d\omega^*y}
\ln^{1/2}\frac{\tilde{u}}{d\omega^*}
e^{-8\sqrt{2}
\tilde{K} \ln^{\frac 1{2}}\frac{\tilde{u}}{(\omega^* y d)}}
\end{equation}
 From (\ref{highconq}) one sees that,
as far as exponents are concerned,
the conductivity still decays as $1/\omega^4$. This would correspond to
a nearly classical ($K \sim 0$) system with short-range interactions.
Note that in this limit the $4 k_F$ scattering is indeed dominant over
the $2 k_F$ one since the latter would only give a conductivity in
$1/\omega^3$ for $K\to 0$ (in the above power laws the frequency is
normalized by the bandwidth so that $\omega \ll 1$).

\section{Temperature dependence of the conductivity}
\label{temperature}

One can use arguments similar to the one introduced in
section~\ref{conductivity} to obtain the temperature dependence of the
conductivity. Instead of having a cutoff length imposed by the frequency
$\omega \sim u/l_\omega$, or more precisely as in (\ref{pinningob})
when Coulomb interactions dominate, one can introduce a thermal length
$\xi_T$ such that $T \sim u/\xi_T$, which will act as a similar cutoff.
Instead of rederiving all the expressions as a function of the
temperature, it is simpler to use the following relation for the
conductivity
\cite{gotze_fonction_memoire,giamarchi_umklapp_1d,giamarchi_attract_1d}
\begin{eqnarray} \label{functional}
\sigma(\omega,T=0) \sim M(\omega,T=0)/\omega^2 \\
\rho(\omega=0,T) \sim M(\omega=0,T) \nonumber
\end{eqnarray}
where $M$ is the so-called memory function. $M$ has the same functional
form depending on the lowest cutoff in the problem. Therefore
$M(\omega,T=0)$ and $M(\omega=0,T)$ have identical form provided one
replaces $\omega$ by $T$. From (\ref{functional}), one sees that it is
possible to obtain the temperature dependence of the resistivity by
multiplying
the frequency form obtained in section~\ref{conductivity}, and then
substituting $\omega$ by $T$. Such a procedure will be valid as long as
one can have a perturbation expansion in the scattering potential, so
that (\ref{functional}) is valid
\cite{giamarchi_umklapp_1d,giamarchi_attract_1d}. This will be the case
as long as the thermal length $\xi_T$ is smaller than the pinning length
$L_0$. Let us examine the various regimes

\subsection{$\xi_T \ll \xi_{cr}$}

As discussed in
section~\ref{weak}, for quantum wires with unscreened long-range Coulomb
interactions, in such a regime a one-dimensional model is probably not
applicable. However, it can have
application either if the long-range interactions are screened or if the
short-range interactions are strong enough ($K$ small) so that
$\xi_{cr} \gg d$.
In that case, as discussed in section~\ref{LargeFreq}, the short-range
interactions dominate. One is back to the situation of $2 k_F$
scattering in a Luttinger liquid for which the temperature dependence of
the conductivity was computed in reference
\cite{giamarchi_loc_lettre,giamarchi_loc}. Let us briefly recall the
results (for a complete discussion see
\onlinecite{giamarchi_loc_lettre,giamarchi_loc}): for repulsive
interactions the conductivity is roughly given by
\begin{equation} \label{rough}
\sigma(T) \sim T^{\frac52 - K(T) - \frac32 K_\sigma(T)}
\end{equation}
where $K(T)$ and $K_\sigma(T)$ are the renormalized Luttinger liquid
parameters for charge and spin at the length scale $\xi_T$. The $K$ are
renormalized by the disorder and decrease when the temperature is
lowered. Such a decrease of the exponents is a signature of the tendency
of the system to localize \cite{giamarchi_loc}. As a
result the conductivity has no simple power law form since the exponents
themselves depend on the temperature. If the disorder is weak enough so
that one can neglect the renormalization of the exponents, one gets the
approximate expression for the conductivity
\cite{apel_impurity_1d,giamarchi_loc} (see
also appendix~\ref{HighFreq} for a rederivation of this result using
SCHA)
\begin{equation} \label{appcond}
\sigma(T) \sim T^{1 - K}
\end{equation}
since (in the absence of renormalization by disorder)  $K_\sigma = 1$
due to spin symmetry. The expression (\ref{appcond}) coincides with the
one obtained subsequently for the conductance of a single impurity
\cite{kane_qwires_tunnel_lettre,kane_qwires_tunnel}. For one single
impurity there
is no renormalization of the exponents \cite{kane_qwires_tunnel} and the
conductance is given by
\begin{equation} \label{kane}
G_0 \sim T^{1-K}
\end{equation}
at all temperatures. If one assumes that there are $N_i$ impurities in
a wire of length $L$ and that the impurities act as {\bf independent}
scatterers, then the conductivity would be, if $G$ is the
conductance of the wire
\begin{equation} \label{gsig}
\sigma(T) = L G = \frac{L}{N_i} G_0 = \frac1{n_i} G_0
\end{equation}
and one recovers (\ref{appcond}) (the impurity density $n_i$ is
included in the disorder in (\ref{appcond})). When many impurities are
present
the assumption that their contributions can be added independently is of
course incorrect. The collective effects of many impurities leads to the
renormalization of the Luttinger liquid parameters (and in particular
to localization) and to the formula (\ref{rough}) for the conductivity
instead of (\ref{appcond}).

\subsection{$\xi_{cr} \ll \xi_T \ll L_0$}

In this regime Coulomb interactions dominate and the $4 k_F$ scattering
is the dominant process. Using (\ref{highconq}) one gets
\begin{equation} \label{intert}
\rho(T) \sim \frac1{T^2}
\ln^{-1/2}\frac{\tilde{u}}{d T} e^{-8\sqrt{2}
\tilde{K} \ln^{\frac 1{2}}\frac{\tilde{u}}{(T d)}}
\end{equation}
Provided the wire is long enough (\ref{intert}) gives also the
temperature dependence of the conductance of the wire.
In this regime the $2 k_F$ scattering would give $\rho_{2 k_F}(T)
\sim 1/T$ and is subdominant. Due to the long-range interactions the
renormalization of the exponents of the conductivity that took place for
short-range interactions \cite{giamarchi_loc} does not take place. Such
a change of exponent with temperature is replaced by sub-leading
corrections. This is due to the fact that the correlation functions
decay much more slowly than a power-law.

\subsection{$L_0 \ll \xi_T$}

This is the asymptotic regime for which the system is pinned and no
expansion like (\ref{functional}) is available. In this regime the
temperature dependence is much less clear. In analogy with the
collective pinning of vortex lattices
\cite{feigelman_collective,nattermann_pinning}, one could
expect a glassy-type nonlinear $I-V$ characteristic of the form
\begin{equation} \label{nonlin}
I \sim e^{-\beta (1/E)^\mu}
\end{equation}
Such an $I-V$ characteristic would correspond to diverging barrier
between metastable states as the voltage goes to zero. (\ref{nonlin})
implies that the linear conductivity vanishes at a finite
temperature. Since this could be viewed as a phase transition (with the
linear conductivity as an order parameter), it is forbidden in a
strictly one dimensional system. In fact, in
a purely one dimensional system (in principle for $d<2$
\cite{nattermann_pinning}), the barriers
should remain finite. In that case one gets a finite linear
conductivity, going to zero when $T\to 0$. A possible form being
\begin{equation}
\sigma(T) \sim e^{-E_B/T}
\end{equation}
where $E_B \sim 1/L_0$ is a typical energy scale for the barriers.
However, no
definite theoretical method exists to decide the issue, and an
experimental determination of the low temperature conductivity would
prove extremely interesting.

\section{Discussion and conclusions}
\label{conclusion}

We have looked in this paper at the conductivity of a one-dimensional
electron gas in the presence of both disorder and long-range Coulomb
interactions. Due to long-range interactions, the electron gas forms a
Wigner crystal which will be pinned by impurities. As a result,
conversely to what happens in a Luttinger liquid, the dominant
scattering corresponds to $4 k_F$ scattering on the impurities and not
$2 k_F$ scattering. Such a pinned Wigner crystal
is close to classical charge density waves but important a
priori differences lie in the presence of long-range interactions and
non-negligible quantum fluctuations.

We have computed the
pinning length above which the (quasi) long-range crystalline order is
destroyed by the disorder. Compared to the standard CDW case the pinning
length is increased both by the Coulomb interactions that makes the
system more rigid and therefore more difficult to pin, and by the
quantum fluctuations that make the pinning less effective. These effects
make the system more likely to be in the weak pinning regime.
We have also computed
the frequency dependence of the conductivity of such a
system. At low frequencies, the conductivity varies as
$\omega^2$ if the pinning on impurities is weak. This is to be
contrasted to the result of
Efros and Shklovskii \cite{shklovskii_conductivity_coulomb}
$\sigma\sim \omega$. We believe that this difference is due to the fact that
their result was derived in a different physical limit, namely when the
pinning length is much shorter than the interparticle distance. However
since the method we use is approximate, it could also be the consequence
of having neglected soliton-like excitations of the phase field.
Although we do expect such excitations to play little role at least when
the short-range repulsion is strong enough ($K$ small). More
theoretical, and especially more experimental investigations would prove
extremely interesting to settle this important issue.
For the case of
strong pinning there is a pseudo-gap in the optical conductivity up to
the pinning frequency. In the pseudo gap
the conductivity behaves as
$\frac1{\omega^2}\ln^{1/2}\frac1{\omega}e^{-1/\omega\ln^{1/2}\frac1{\omega}}$.
Above the pinning
frequency, for both regimes,
the conductivity decreases as $1/\omega^4\ln^{-1/2}(\omega_{cr}/\omega)
e^{-\text{Cste}\ln^{1/2}(\omega_{cr}/\omega)}$
up to the crossover frequency $\omega_{cr}$ above which
the long-range Coulomb interactions become unimportant.
For the parameter we took here, $\omega_{cr}$ is also the limit of
applicability of a one-dimensional system since $\xi_{cr} \sim d$, the
width of the wire. However if the short-range interactions are strong
enough ($K$ small), so that $\xi_{cr} \gg d$, then above $\omega_{cr}$
a one-dimensional description will still be valid.
One is back to the situation of $2 k_F$ scattering in a Luttinger
liquid which was studied in detail in \onlinecite{giamarchi_loc}.
The conductivity then behaves as
$(1/\omega)^{\mu(\omega)}$, where $\mu(\omega)$ is a non universal
exponent depending on the short-range part of the interactions, and due
to the renormalization of Luttinger liquid parameters by disorder, also
dependent on the frequency \cite{giamarchi_loc}. If one can neglect such
a renormalization of the exponents (e.g. for very weak disorder) then
$\mu = 3 - K$.

The temperature dependence can be obtained by similar methods.
One can
define a thermal length $\xi_T\sim u/T$.
When $\xi_T < L_0$, the
frequency and temperature dependence of the conductivity are simply
related by $\rho(\omega=0,T) \sim T^2 \sigma(\omega\to T, T=0)$, giving
$\rho(T) \sim 1/T^2\ln^{-1/2}(1/T) e^{-\text{Cste}\ln^{1/2}(1/T)}$.
Above the pinning length, frequency and temperature
can no longer be treated as equivalent cutoffs, and the conductivity is
much more difficult to compute.
On can expect an exponentially vanishing linear conductivity, provided
that the barriers between metastable states remain finite. If it is not
the case, one should get a non-linear characteristic of the form $I \sim
\text{exp}[-\beta (1/E)^\alpha]$, where $\beta =1/T$, and $\alpha$ is an
exponent. Again an experimental determination of $\sigma(T)$
would prove extremely useful. Note that
although we considered here the conductivity, most of the results can be
applied to the conductance of a finite wire, provided that the size of
the wire $L$ is larger than the thermal length $L_T$.

We know that under the application of strong enough electric fields,
a classical CDW can be depinned \cite{littlewood_sliding_cdw,lee_depinning}.
Similarly we expect for a
Wigner crystal the existence of a threshold electric field $E_{th}$
above which a finite static conductivity appears.
We can  make a crude estimation of this threshold field,
made on the simple assertion that the electrical energy at threshold must
be of the order of the pinning energy
$\omega^* \sim \frac{\tilde{u}}{L_0}\ln^{1/2}\frac{L_0}{d}$.
This energy
can be written as $eU$, where $U$ is the electrical potential corresponding
to a segment of the wire of length $L_0$, that is to say $U=E_{th}L_0$.
Thus the threshold field is estimated as
\begin{equation}
\label{threshold}
E_{th} \sim \frac{\tilde{u}}{eL_0^2}\ln^{1/2}\frac{L_0}{d}
\end{equation}
 From this we can extract an estimation of the pinning frequency $\omega^*$.
Indeed  experimental values of such threshold fields can be found in the
literature\cite{kastner_coulomb}. The latter reference gives a value of
threshold field, for a wire of length of about $10\mu m$, of
$E_{th}=5\times 10^2.V.m^{-1}$. Thus (\ref{threshold}) gives
$L_0 \sim 1.4 \mu m$, which seems quite reasonable compared to the
length of the wire, and gives for the pinning frequency the estimation
$\omega^* \sim 1\times 10^{12}Hz$ (since the wires
of reference \onlinecite{kastner_coulomb} contain typically two or three
impurities, one is probably here in a strong pinning regime).
Data reported here are just meant as typical values. The system
studied in \onlinecite{kastner_coulomb} is at the limit of
applicability of our study at low temperatures,
since the wire is so short that it contains only few impurities. However,
regardless of the pinning mechanism and number of impurities,
our theory should give correctly the
temperature dependence of the conductivity (or conductance) at
temperatures such that $\xi_T < n_i^{-1}$, since in this regime the
impurities act as independent scatterers. To make a study of the low
temperature/low frequencies properties longer wires would be needed.

The above estimates, although very crude, show that
typical frequencies or
temperatures for such systems are in the range of experimentally realizable
values, which gives hope for more experimental evidence for the
existence of such pinned Wigner crystals. In particular, measurements
of the temperature dependence of the conductivity/conductance would
prove decisive. At low temperature they would provide evidence for a
pinned Wigner crystal, and at higher temperature test for the
scattering on impurities in the presence of long-range interactions
($\sim 1/T^2$ behavior). A possible crossover between a Luttinger liquid
(dominated by short-range interactions) and the
Wigner Crystal (dominated by long-range interactions) could also in
principle be seen on the temperature dependence of the conductivity.
Frequency dependent conductivity measurement are probably much more
difficult to carry out, but would be also of high importance. At low
frequency they could serve as tests both on the nature of the pinning
mechanism and on the effects on long-range Coulomb interactions on the
frequency dependence of the conductivity. For these purposes, quantum
wires would constitute a much cleaner system than the standard CDW
compounds.

\acknowledgments

We would like to thank H.J. Schulz for many stimulating discussions. One
of us (T.G.) would like to thank H. Fukuyama, M. Ogata,
B. Spivak and V. Vinokur for interesting discussions,
and the Aspen center for Physics where part of this work was completed.

\appendix

\section{Effect of quantum fluctuations at low frequencies}
\label{quant}

In this section we calculate the corrections to the pinning length $L_0$
(in the weak pinning case)
introduced by the quantum effect, at zero temperature.
We shall use the SCHA method to take it into account \cite{suzumura_scha}.
We write
\begin{equation}
\Phi =\Phi_{cl}+\hat{\Phi}
\end{equation}
where $\Phi_{cl}$ is the classical solution studied in section \ref{static}
and $\hat{\Phi}$ represents
the quantum fluctuations around $\Phi_{cl}$. We then use the relation
\begin{equation}
\cos (Qx+\Phi_{cl}+2\sqrt{2}\hat{\Phi})=e^{-4\langle \hat{\Phi}^2(x)
\rangle}
\cos(Qx+2\sqrt{2}\Phi_{cl}) (1-4(\hat{\Phi}^2(x)-\langle \hat{\Phi}^2(x)
\rangle))
\end{equation}
instead of the classical expansion done in (\ref{expansion}).

Thus the hamiltonian $H_{\hat{\Phi}}$ in terms of $\hat{\Phi}$ is
similar to (\ref{expansion}), but has an additional constant part
\begin{equation}
H'_{pot}=\sum_j V_0\rho_0e^{-4\langle \hat{\Phi}^2(x) \rangle}
\cos (QX_j+2\sqrt{2}\Phi_{cl})
(1+4\langle \hat{\Phi}^2(x) \rangle)
\end{equation}

We note $\gamma \equiv e^{-4\langle \hat{\Phi}^2(x) \rangle}$ and
$\langle \hat{\Phi}^2 \rangle=\langle \hat{\Phi}^2(x) \rangle$.
 $\langle \hat{\Phi}^2 \rangle$ has to be determined self-consistently from
\begin{equation} \label{fluct}
\langle \hat{\Phi}^2 \rangle=
{K \over 2}\int_0^{\Lambda^{-1}} dq
{1\over \sqrt{q^2(1+{2KV(q) \over \pi u}) + q_c^2}}
\end{equation}
where $\Lambda^{-1}$ is a cut-off the choice of which is discussed in the
following and
\begin{equation} \label{correction}
{1 \over 2\pi}{u \over K}q_c^2=-4V_0\rho_0
e^{-4\langle \hat{\Phi}^2 \rangle}
{1 \over L}\langle \sum_j \cos (QX_j+2\sqrt{2}\Phi_{cl})\rangle_{av}
=4V_0\rho_0({n_i \over L_0})^{1/2}\gamma \equiv 4
\overline{V}\gamma
\end{equation}
Indeed the potential of impurities results in a change of the spectrum
$\omega (q)$:
\begin{equation}
\omega (q)_{V_0 \ne 0} = u\sqrt{q^2 (1+{2KV(q) \over \pi u}) +q_c^2}
\approx u\sqrt{q^2 (1-{4Ke^2 \over \pi u}\ln (qd)) +q_c^2}
\end{equation}

We point out that (\ref{correction}) gives
$uq_c \sim \frac{\tilde{u}}{L_0} \sim \omega^*$, where
$\tilde{u}=u\sqrt{\frac{4Ke^2}{\pi u}}$ and $\omega^*$ is the characteristic
frequency of the system, defined in section \ref{weak} by (\ref{PinFreq}),
where we show that $\omega^* \ll \frac{u}{d}$.
Consequently $q_cd \sim \omega^* \frac d{u} \ll 1$.
Thus we can choose $\Lambda^{-1} > q_c$ such that
$\alpha_cK_0(qd)\approx -\alpha_c\ln(qd)$ in the whole range
 of integration. Furthermore $\ln\frac1{q_cd} \sim 5$ is large compared to
unity.
Using those remarks we found
\begin{equation}
\langle \hat{\Phi}^2 \rangle \approx
\tilde{K} \ln^{1/2} \frac{\alpha_c}{q_c^2d^2}
\end{equation}
where $\tilde{K} = {\sqrt{\pi uK} \over 2\sqrt{2}e}$.

Replacing $q_c$ by its value, we find a square equation on
$\langle \hat{\Phi}^2 \rangle$. The result is simplified using
$ln{1 \over q_cd} \gg 1$ .Thus we find
\begin{equation}
\langle \hat{\Phi}^2 \rangle \approx \tilde{K}
\ln^{1/2} \frac {e^2}{d^2\overline{V}}
\end{equation}
Now we can calculate $L_0$ by estimating the cost in
energy of a variation of the phase $\Phi_{cl}$ due to the
impurity potential $V_0\rho_0$.
The costs in elastic and Coulomb energy are again given by
(\ref{eps-el}) and (\ref{eps-coul}) respectively. The cost in potential energy
slightly differs from (\ref{eps-imp}):
\begin{equation} \label{new-imp}
{\cal E}_{\rm imp}(L_0)=- \overline{V}\gamma
(1+4\langle \hat{\Phi}^2 \rangle)
\end{equation}
The new contribution we have to add is the variation of the zero-point energy
due to the fact that the spectrum  is modified at $V_0 \ne 0$. It is, by
unit length
\begin{eqnarray}
\delta {\cal E} & = & {1 \over L}{1 \over 2} \sum_q
\omega (q)_{V_0 \ne 0} -\omega (q)_{V_0 = 0} \nonumber \\
 & = & u \int_0^{\Lambda^{-1}} {dq \over 2\pi}
\sqrt{q_c^2+ q^2(1+{4Ke^2 \over \pi u}K_0(qd)) }
- \sqrt{ q^2(1+{4Ke^2 \over \pi u}K_0(qd)) }
\end{eqnarray}
We can approximate
\begin{equation}
\delta {\cal E}\approx
u \int_0^{\Lambda^{-1}} {dq \over 2\pi}
\sqrt{q_c^2- {4Ke^2 \over \pi u}q^2\ln qd }-\sqrt{-
{4Ke^2 \over \pi u}q^2\ln qd }
\end{equation}
In a first step we find
\begin{equation}
 \delta {\cal E}={1 \over 2}{uq_c^2 \over 2\pi}
\int_0^{\Lambda^{-1}} dq {1\over \sqrt{q_c^2-{4Ke^2 \over \pi u}q^2\ln qd }}
+{1 \over 4}{uq_c^2 \over 2\pi}
\sqrt{\pi u \over 4Ke^2}\ln^{-1/2}{\Lambda \over d}
\end{equation}
Using (\ref{fluct}) it is rewritten\,:
\begin{equation} \label{zero-point}
 \delta {\cal E}= \overline{V}\gamma
(4\langle \hat{\Phi}^2 \rangle +\sqrt{2}\tilde{K}
\ln^{-1/2}{\Lambda \over d})
\end{equation}
We remark that the terms in $\langle \hat{\Phi}^2 \rangle$ annihilate when
summing (\ref{new-imp}) and (\ref{zero-point}).
\begin{equation}
{\cal E}_{\rm imp}+\delta {\cal E}=
-V_0\rho _0 ({n_i \over L_0})^{1 \over 2}
e^{-4\langle \hat{\Phi}^2 \rangle}
(1-\sqrt{2}\tilde{K}\ln^{-1/2}{\Lambda \over d})
\approx -\overline{V}\gamma
\end{equation}
The previous approximation follows from $\ln \frac{\Lambda}{d} \gg 1$.
In the case of short-range interactions
\cite{suzumura_scha}
$(1-\sqrt{2}\tilde{K}\ln^{-1/2}{\Lambda \over d})$ would be
replaced by $(1-2K)$ and one could not neglect that correction.
Here the calculation finally reduces to the replacement of $V_0\rho _0$ by
$V_0\rho _0e^{-4\langle \hat{\Phi}^2 \rangle}=V_0\rho _0\gamma$,
which does not change the exponents since
$\gamma \approx e^{-4\tilde{K}\ln ^{1/2} \frac1{\overline{V}}}$.

We find instead of (\ref{length})
\begin{equation}
L_0 =({8e^2 \over \alpha \pi ^2 V_0\rho _0\gamma n_i^{1 \over 2}})^{2 \over 3}
\ln ^{2 \over 3} ({CL_0 \over d})
\end{equation}
where $\gamma \approx
e^{-\frac{8\tilde{K}}{\sqrt{3}}\ln ^{1/2} \frac1{V_0\rho _0}}$.

\section{Effect of quantum fluctuations at high frequencies}
\label{HighFreq}

As explained in (\ref{LargeFreq}), we handle the quantum fluctuations at
high frequencies by
a method analogous to SCHA but where we assume that the system is cut
into subparts of length $l_\omega \sim \frac{\tilde{u}}{\omega}$.
Such a description is expected to be valid as long as $l_\omega <
L_0$. This provides, instead of the effective
cut-off $q_c$ due to disorder as in (\ref{fluct}), a natural
cut-off $\frac \omega{\tilde{u}}$ which appears at frequencies
$\omega > \tilde{u} q_c$. Thus for high frequencies we use, instead of
(\ref{fluct})
\begin{equation}
\langle \hat{\Phi}^2 \rangle=
\frac K{2}\int_{\frac \omega{\tilde{u}}}^{\Lambda^{-1}} dq
\frac1{\sqrt{q^2(1+{2KV(q) \over \pi u}) + q_c^2}}
\approx
\frac K{2}\int_{\frac \omega{\tilde{u}}}^{\Lambda^{-1}} dq
\frac1{\sqrt { q^2(1+\frac{2KV(q)}{\pi u}) }}
\end{equation}

In the first subsection we make that calculation for our problem, at
$\omega \ll \omega_{cr}$ and in the second one we derive as a comparison
what the corrections would be in the absence of the Coulomb term.
In the following we can take $\Lambda=d$ since we stay in the one-dimensional
regime $\omega \ll \frac{u}{d}$.

\subsection{long-range interactions}

In the range $\omega^* \ll \omega \ll \omega_{cr}$ we have
\begin{equation}
\langle \hat{\Phi}^2 \rangle
\approx
\frac K{2}\int_{\frac \omega{\tilde{u}}}^{d^{-1}} dq
\frac1{ q\sqrt{\frac{2KV(q)}{\pi u} }}
\approx
\sqrt{2}\tilde{K} \ln^{\frac 1{2}}\frac{\tilde{u}}{(\omega d)}
\end{equation}
Since we are in the limit where $l_\omega < L_0$, and are dealing with
segments of size $l_\omega$, we can replace the pinning length by
$l_\omega$ in the calculation of the conductivity, computed
using $\Sigma \approx \Sigma_1+\Sigma_2$.
We have
\begin{eqnarray} \label{sigsig1}
\Sigma_1 &=& - 8 V_0 \rho_0 (\frac{n_i}{l_\omega})^{\frac1{2}}
e^{-4\langle \hat{\Phi}^2 \rangle}
\text{ for } l_\omega\ge n_i^{-1} \nonumber \\
&=& - 8 V_0 \rho_0n_ie^{-4\langle \hat{\Phi}^2 \rangle}
 \text{ for } l_\omega\le n_i^{-1}
\end{eqnarray}
and
\begin{equation}  \label{sigsig2}
\Sigma_2 = c_\Phi n_i (8 V_0 \rho_0 \gamma)^2
\int_{\frac{|\omega_n|}{u}}^{d^{-1}}
\frac{dq}{2\pi}\frac{\pi uK}{\omega _n^2 + q^2u^2(1+{2KV(q) \over \pi u})}
\end{equation}
where $c_\Phi$ is defined in (\ref{cphi}).
Since $\Sigma_2$ is multiplied by $\gamma^2$ compared to the classical
case, the conductivity becomes instead of (\ref{highcon})
\begin{equation}
\omega^* \Re e \sigma (y) = \frac{c_{\Phi} 4uKe^2}{\pi}
y^{-4}\ln^{-1/2}\frac{\tilde{u}}{d\omega^*y}
 \ln^{1/2}\frac{\tilde{u}}{d\omega^*}e^{-8\sqrt{2}
\tilde{K} \ln^{\frac 1{2}}\frac{\tilde{u}}{(\omega^* y d)}}
\end{equation}

\subsection{short-range interactions}

Here we rederive the results for short-range interactions obtained by
the renormalization group \cite{giamarchi_loc_lettre,giamarchi_loc}.
As pointed out in \onlinecite{giamarchi_loc}, the SCHA does not
correctly reproduces the renormalization of the exponents with disorder
and is therefore limited, for the case of short-range interactions, to
describe infinitesimal disorder. For finite disorder the exponent
themselves are renormalized by the disorder and are functions of
frequency or temperature \cite{giamarchi_loc_lettre,giamarchi_loc}.
The derivation is given for simplicity for the case of a $4 k_F$
scattering. In the case of short-range interactions one has to consider
the $2 k_F$ scattering (which is dominant). Since one has
\begin{eqnarray}
\rho_{4 k_F}(x) & \sim & e^{i 2 \sqrt2 \Phi(x)} \\
\rho_{2 k_F}(x) & \sim & e^{i \sqrt2 \Phi(x)} \cos(\sqrt2 \Phi_\sigma(x))
\nonumber
\end{eqnarray}
where $\Phi$ and $\Phi_\sigma$ are two independent bosons for charge
and spin excitations, one can obtain the conductivity for the $2 k_F$
scattering by the substitution $4 K \to K + 1$ (see
\onlinecite{giamarchi_loc}
for a complete derivation of the conductivity for $2 k_F$ scattering).

For short-range interactions
\begin{equation}
\langle \hat{\Phi}^2 \rangle=
\frac K{2}\int_{\frac \omega{u}}^{d^{-1}} \frac {dq}{q}
= \frac K{2} \ln \frac{ud^{-1}}{\omega}
\end{equation}
which gives
\begin{equation}
\gamma = e^{-4\langle \hat{\Phi}^2 \rangle}
=\Bigl(\frac{ud^{-1}}{\omega}\Bigr)^{-2K}
\end{equation}
We have
\begin{eqnarray}
\Sigma_1 &=& - 8V_0\rho_0(\frac{n_i}{l_\omega})^{\frac1{2}}
e^{-4\langle \hat{\Phi}^2 \rangle}
\sim - (\frac{\omega d}{u})^{2K+\frac12}
\text{ for } l_\omega \ge n_i^{-1} \nonumber \\
&=& - 8V_0\rho_0n_ie^{-4\langle \hat{\Phi}^2 \rangle}
\sim - (\frac{\omega d}{u})^{2K}\text{ for } l_\omega \le n_i^{-1}
\label{sigmasat}
\end{eqnarray}
and
\begin{equation}
\Sigma_2=c_{\Phi}n_i (8V_0\rho_0\gamma)^2
\int_{\frac{|\omega_n|}{u}}^{d^{-1}}
\frac{dq}{2\pi}\frac{\pi uK}{\omega_n^2+q^2u^2}
\end{equation}
The previous integral equals a constant times $\frac1{|\omega_n|}$.
Taking the limit $i\omega_n \to \omega-i0^+$,
\begin{equation}
\Sigma_2 \sim -i(\frac{\omega d}{u})^{4K}(\omega)^{-1}
\end{equation}
We now compute the frequency dependence of the conductivity.
 From (\ref{sigmasat}), one sees that $\Sigma_1 \ll \omega^2$  at high
frequencies.
 From (\ref{conduct}) one finds
\begin{equation}
{\cal R} e \sigma_{4 k_F}
(\omega) \sim (\frac{\omega d}{u})^{4K}\omega^{-4}
\end{equation}
Similarly one would get $\sigma_{2 k_F}(\omega) \sim \omega^{K-3}$


\newpage
\begin{figure}
\caption{\label{conducti}
Frequency dependence of the rescaled conductivity in the weak pinning case
versus the rescaled frequency $y=\frac\omega{\omega^*}$.
We have noted $\zeta = \frac{2uKe^2}{\pi \omega^*}$.
The solid line was obtained by numerical resolution of equation
(\protect{\ref{exacte}}),
where we have taken typical parameters $\frac{\omega^*d}u = 3.33\times10^{-2}$
and $\frac{4Ke^2}{\pi u}=4.7$.
The dash-dotted curve represents the analytical expression
(\protect{\ref{result}}) computed for low frequencies.}
\end{figure}
\begin{figure}
\caption{\label{conductiStrong}
Frequency dependence of the rescaled conductivity in the strong pinning case
versus the rescaled frequency $y=\frac\omega{\omega_0}$.
We have noted $\zeta = \frac{2uKe^2}{\pi \omega_0}$.
The three curves were obtained by solving numerically
equation (\protect{\ref{self}})
for different values of the strength of the pinning
$\epsilon$: $\epsilon=1$ (dashed line), $\epsilon=10$ (dash-dotted line), and
$\epsilon=1000$ (solid line). We have taken
$\frac{\omega_0d}u = 3.33\times10^{-2}$ and $\frac{4Ke^2}{\pi u}=4.7$.}
\end{figure}
\end{document}